\documentclass[aps,prb,twocolumn,superscriptaddress,floatfix,showpacs]{revtex4}

\usepackage{graphicx}
\usepackage{amsmath}
\usepackage{color}

\def\c{\hat{c}^{}}
\def\cc{\hat{c}^{+}}

\def\mz2{\langle m_z^2 \rangle}
\def\eps{\varepsilon}
\def\s{\sigma}

\def\w{i\omega_n}
\def\bk{\mathbf{k}}

\def\etal{\textit{et al.}}

\def\invar{Fe$_{1-x}$Ni$_x$}

\begin{document}
\title{Magnetic properties of \invar~alloy from CPA+DMFT perspectives}

\author{Alexander I. Poteryaev}
\affiliation{M. N. Miheev Institute of Metal Physics of Ural Branch of Russian Academy of Sciences, 18, S. Kovalevskaya Str., 620990 Ekaterinburg, Russia}
\author{Nikolay A. Skorikov}
\affiliation{M. N. Miheev Institute of Metal Physics of Ural Branch of Russian Academy of Sciences, 18, S. Kovalevskaya Str., 620990 Ekaterinburg, Russia}
\author{Vladimir I. Anisimov}
\affiliation{M. N. Miheev Institute of Metal Physics of Ural Branch of Russian Academy of Sciences, 18, S. Kovalevskaya Str., 620990 Ekaterinburg, Russia}
\author{Michael A. Korotin}
\affiliation{M. N. Miheev Institute of Metal Physics of Ural Branch of Russian Academy of Sciences, 18, S. Kovalevskaya Str., 620990 Ekaterinburg, Russia}

\begin{abstract}
We use a combination of the coherent potential approximation and dynamical mean field theory
to study magnetic properties of the \invar~alloy from a first principles. 
Calculated uniform magnetic susceptibilities have a Curie-Weiss-like behavior and extracted
effective temperatures are in agreement with the experimental results.
The individual squared magnetic moments obtained as function of nickel concentration
follow the same trends as experimental data.
An analysis of the ionic and spin weights shows a possibility of a high-spin to intermediate- and
low-spin states transitions at high temperatures.
\end{abstract}

\pacs{71.20.Be, 71.23.-k, 71.27.+a}


\maketitle

\section{Introduction\label{sec:intro}}

Materials with transition metal elements are of great interests due to a large variety
of physical properties: high-temperature superconductivity in copper and iron based compounds~\cite{highT},
giant magnetoresistance~\cite{Fert2008}, metal-to-insulator transition~\cite{Imada1998} and many others.
This diversity comes mainly from the facts that {\em i)} the local Coulomb interaction of the 3$d$ ions 
is comparable with its bandwidth and {\em ii)} the $d$ orbitals are partially filled.
These compounds with competing kinetic and interacting energies can hardly be described
by the conventional density functional theory (DFT) and are known as strongly correlated.

The elemental 3$d$ metals with partially occupied shell are not exceptions.
The famous 6~eV satellite in nickel~\cite{Guillot1977} is of many body nature and cannot
be described by a band structure theory~\cite{Clauberg1981}.
An application of the state of the art DFT+DMFT method~\cite{LDA+DMFT}, that combines material specific aspects
of the DFT and dynamical mean field theory~\cite{Georges1996} (DMFT) to treat correlated electrons,
allowed to Lichtenstein \etal~\cite{Lichtenstein2001} describe properly electronic and magnetic properties of 
iron and nickel in ferro- and paramagnetic phases.
Grechnev \etal~\cite{Grechnev2007}, Di Marco \etal~\cite{DiMarco2009}, 
Koloren\v{c} \etal~\cite{Kolorenc2012} and Min\'{a}r \etal~\cite{Minar2013} studied spectral properties of Ni and 
found that correlation effects play 
an important role in forming satellite structure and renormalize the exchange splitting.
Benea \etal~\cite{Benea2012} calculated magnetic Compton profiles of Ni and Fe 
in DFT+DMFT approach and found that an inclusion of electronic correlations 
improves significantly an agreement between the theoretical and experimental results.
Leonov \etal~in series of papers~\cite{Leonov2011,Leonov2012,Leonov2014} investigated
a paramagnetic iron at ambient pressure as function of temperature and 
found that a complete picture of phase transformations can be done 
when correlation effects are taken into account. 
Pourovskii \etal~\cite{Pourovskii2013,Pourovskii2014,Igoshev2013} studied 
magnetic and thermodynamic properties of iron at high pressure and concluded that correlations
are important for phase stability.

In this view, an account for strong electron-electron interaction is required 
to investigate physical properties of \invar~alloy, which was intensively studied 
due to the Invar effect discovered more than hundred years ago~\cite{Guillaume1897}.
One should note that a chemical disorder present in alloy complicates additionally
a problem of a first principles description of materials.
Nevertheless, different techniques to treat alloying effects are well developed
in framework of DFT. 
In one of the most used methods, a large supercell with randomly distributed atoms of different types is constructed,
and then, 
required properties can be evaluated from calculations for ordered structures.
%
%
In a coherent potential approximation (CPA), a real crystal with randomly distributed ions
is replaced by an effective medium with an energy-dependent self-energy that has to be determined
self-consistently (for recent reviews see~[\onlinecite{Ruban2008}] or [\onlinecite{Rowlands2009}] and references therein).

Magnetic properties of \invar~alloy as function of composition and
volume were studied by several groups using both above mentioned approaches in the DFT framework.
A rather complete review of these works can be found in paper of Abrikosov \etal~\cite{Abrikosov2007}
who also found that a family of magnetic states are close to each other in energy and
concluded that inclusion of strong electron-electron correlations is highly desired
to reveal the physics of this alloy. 
Albeit the pure iron and nickel are successfully studied by means of DFT+DMFT method,
there are only few papers devoted to \invar.
Min\'{a}r \etal~\cite{Minar2005,Minar2014} combined a Korringa-Kohn-Rostoker multiple scattering theory 
with the dynamical mean field theory and coherent potential approximation to treat substitutional disorder.
For a solution of the effective impurity problem in DMFT they used fluctuation exchange approximation~\cite{Bickers1991} 
in conjunction with $T$-matrix theory~\cite{T-matrix}.
The authors found that a band broadening due to correlation effects and disorder is comparable
in face centered cubic structure (fcc) of \invar~below Curie temperature.
Total magnetic moment is decreased as function of Ni concentration.
The individual magnetic moments for iron (nickel) are slightly overestimated (underestimated)~\cite{Minar2014}
with respect to experimental values that can be connected to the perturbation nature of the impurity solver
in DMFT.  
Vekilova \etal~\cite{Vekilova2015} investigated different phases of ordered
Fe$_{0.75}$Ni$_{0.25}$ alloy at high pressure using DFT+DMFT method with continuous time quantum Monte-Carlo
as the impurity solver~\cite{Werner2006}. 
For a body centered cubic structure, a uniform magnetic susceptibility has a Curie-Weiss-like behavior,
which changes to a Pauli-type in a hexagonal close packed structure.
They found that 
a strength of electronic correlations on the Fe 3$d$ orbitals is sensitive to the phase and local environment.

In this work, we study the magnetic properties of iron-rich fcc \invar~binary alloy as function of composition
and temperature. 
To this aim, we utilize a combination of the coherent potential approximation with the dynamical mean field theory,
which is described in the next section. 

\section{CPA+DMFT Method\label{sec:method}}

The Hamiltonian of a crystal with a chemical type of disorder can be written as:
\begin{eqnarray} \label{eq:hamiltonian}
 \hat{H} & = &  \sum_{i \neq j} \sum_{m,m',\s} t_{im,jm'}^{\s} \cc_{im\sigma} \c_{jm'\sigma} 
         +  \sum_{i,m,\s} ( \epsilon^i_{m\s} - \mu ) \hat n^i_{m\s}      \nonumber   \\ 
         & + & \hat H_{int},
\end{eqnarray}
where $\cc_{im\s}$ ($\c_{im\s}$) is a creation (annihilation) operator
of an electron on site $i$ with orbital $m$ and spin $\s$.
${\hat n^i_{m\s}=\cc_{im\s} \c_{im\s}}$.
$\mu$ is a chemical potential,
$t_{im,jm'}^{\s}$ are hopping amplitudes,
$\epsilon^i_{m\s}$ is on-site energy.
The last term in the Hamiltonian~(\ref{eq:hamiltonian}) corresponds to the on-site Coulomb interaction:
\begin{equation} \label{eq:Hcoul}
  \hat H_{int} = \frac{1}{2} \sum_{i,\{m\}} \sum_{\s\s'} U^i_{m'_1 m'_2 m_1 m_2}
                 \cc_{m'_1\s} \cc_{m'_2\s'} \c_{m_1\s'} \c_{m_2\s},
\end{equation}
where $U^i_{m'_1 m'_2 m_1 m_2}$ are elements of the Coulomb interaction matrix.
%
For the case of a substitutional disorder 
the on-site potential, $\epsilon^i_{m\s}$, and the Coulomb matrix, $U^i_{m'_1 m'_2 m_1 m_2}$,
depend on site index $i$ and they are different for different atomic species 
$\epsilon^A_m$ ($U^A_{m'_1 m'_2 m_1 m_2}$) or 
$\epsilon^B_m$ ($U^B_{m'_1 m'_2 m_1 m_2}$), depending on atom remaining on site $i$ with the probability 
$x_A$ or $x_B$, $x_A+x_B=1$.
The hopping amplitudes are assumed to be site-independent.
This approximation seems reasonable for constituents with similar electronic structures,
when the on-site local potentials are close in energy relative to their bandwidth~\cite{Faulkner1982,Ducastelle1991,Abrikosov1998}.

The above Hubbard-type Hamiltonian~(\ref{eq:hamiltonian}) cannot be solved directly due to 
a simultaneous presence of the disorder and interaction.
At the same time, the solutions of limiting cases, though approximate, are known.
In the non-interacting limit, a coherent potential approximation~\cite{Soven67} is utilized 
to describe the chemically disordered crystals. 
In the CPA, an electron propagation through the substitutionally
disordered material is replaced by its propagation
through an effective medium defined self-consistently. 
Other limiting case -- absence of disorder (one type of atoms)  -- is known as the Hubbard model.
One of the best single site approximation for its solution is the dynamical mean field theory~\cite{DMFT}.
In this theory the lattice problem is replaced by an effective impurity embedded in an energy dependent
effective medium. The last has to be found self-consistently.
Hence, both limiting cases at the hand share the effective medium or mean field interpretation of the problem.

To deal with the Hamiltonian~(\ref{eq:hamiltonian}) we will use the same effective medium ideology.
Let us assume for the moment that the hybridization, $\Delta(\w)$, of the embedded atom with the effective medium is known.
Then, the local Green's functions are defined as
\begin{equation}
  G_i = \frac{ \int_0^{\beta} \c \cc e^{-S_i} \mathcal{D}\c\mathcal{D}\cc }
             { \int_0^{\beta} e^{-S_i} \mathcal{D}\c\mathcal{D}\cc },
  \label{eq:G_imp}
\end{equation}
where $\c$ and $\cc$ are Grassman variables and an action is given as:
\begin{widetext}
  \begin{eqnarray} \label{eq:action}   
    S_i = - \sum_{nm\s} \cc_{m\s}(\w) [ \mu + \w - \epsilon_{m\s}^i - \Delta(\w) ] \c_{m\s}(\w)  
        + \frac{1}{2}\int_0^{\beta} d\tau \sum_{\{m\}} \sum_{\s\s'} U^i_{m'_1 m'_2 m_1 m_2}
                 \cc_{m'_1\s}(\tau) \cc_{m'_2\s'}(\tau) \c_{m_1\s'}(\tau) \c_{m_2\s}(\tau),
  \end{eqnarray}
\end{widetext}
with $\omega_n = (2n+1)\pi / \beta$ are the fermionic Matsubara frequencies, 
$\beta = 1 / k_B T$ is an inverse temperature and $\tau$ is an imaginary time.
Self-consistency conditions require that the local Green's function should be equal to 
a weighted sum of impurity Green's functions:
\begin{eqnarray} \label{eq:scf_conditions}
  G(i\omega_n) = x_A G_{A}(i\omega_n) + x_B G_{B}(i\omega_n).
\end{eqnarray}
Then, the local $G(i\omega_n)$ from Eq.~(\ref{eq:scf_conditions})
can be used to compute a self-energy from the Dyson's equation:
\begin{eqnarray} \label{eq:sigma}
  \Sigma(\w) = \w - \Delta(\w) - G^{-1}(\w).
\end{eqnarray}
One should emphasize that the obtained self-energy contains effects from the disorder and
correlations simultaneously. 
This self-energy is utilized to calculate the local Green's function of effective 
medium
\begin{eqnarray} \label{eq:gloc}
 G(\w) = \frac{1}{V_\textrm{BZ}} 
         \int_{BZ} \frac{d\bk}{ (\mu+\w) I - H(\bk) - \Sigma(\w)},
\end{eqnarray}
where $H(\bk)$ is the Fourier transform of the first term of the Hamiltonian~(\ref{eq:hamiltonian}) and
an integration is performed over the first Brillouin zone with volume $V_\textrm{BZ}$.
The new hybridization function is then written as:
\begin{eqnarray} \label{eq:delta}
  \Delta(\w) = \w - \Sigma(\w)  - G^{-1}(\w).
\end{eqnarray}

The above set of equations are iteratively solved until a convergence with respect 
to the self-energy or hybridization is achieved.
One should note, that the above described CPA+DMFT scheme
behaves correctly in the different limits. Namely, if there is one type of atomic species 
all equations reduce to the conventional DMFT set. Other limit is non-interacting ($U^\textrm{A}$=$U^\textrm{B}$=0) and
the equations become of classical CPA.

The generalization for a case of first principles calculations is straightforward. 
The DFT+DMFT Hamiltonian reads as 
\begin{equation} \label{eq:dft+dmft}
  \hat{H}_{DFT+DMFT} = \hat{H}_{DFT} + \hat{H}_{int} - \sum_{im\s} \eps^i_{dc} \hat{n}^i_{m\s},
\end{equation}
where $\hat{H}_{DFT}$ is the Hamiltonian in a localized basis set obtained within density functional theory
framework. The second term is the interaction and it is described by the eq.~(\ref{eq:Hcoul}).
The density functional theory treats the Coulomb interaction in an averaged way, and therefore,
one needs to subtract the so-called double counting term. The double counting is on-site quantity and
we use the fully localized limit in our calculations with the next definition for potential
\begin{equation} \label{eq:dc}
  \eps^i_{dc} = \bar{U}^i \left( N^i_d - \frac{1}{2} \right),
\end{equation}
where $\bar{U}^i$ is a mean Coulomb interaction and 
$N^i_d = \sum_{m\s} n^i_{m\s}$ is a total number of $d$ electrons in DFT calculations.

The DFT+DMFT Hamiltonian (\ref{eq:dft+dmft}) can be easily mapped to the disordered Hamiltonian (\ref{eq:hamiltonian}).
In such case a matrix of hopping integrals in the eq.~(\ref{eq:hamiltonian}) become hoppings calculated in DFT, $t=t(DFT)$,
and the on-site energy is replaced as
\begin{equation}
  \epsilon^i_{m\s} = \epsilon^i_{m\s}(DFT) - \eps^i_{dc},
\end{equation}
where the right-hand part of the equation can be completely defined within conventional band structure calculations.

\section{Results\label{sec:results}}

For a first principles calculations, the Hamiltonian and related parameters (see eq.~(\ref{eq:dft+dmft}))
were obtained using a full-potential linearized augmented-plane wave method implemented 
in the Exciting-plus code (a fork of ELK code~\cite{elk}).  
The exchange-correlation potential was chosen in the Perdew-Burke-Ernzerhof form~\cite{PBE} of 
the generalized gradient approximation. 
The non-spin polarized Hamiltonian, $H(\bk)$, for fcc iron was constructed 
in a basis of well-localized Wannier functions~\cite{proj_wf} containing $s$, $p$ and $d$ states.
A reciprocal space was divided
on the 18$\times$18$\times$18 $\textbf{k}$-points mesh in the full Brillouin zone.
We utilized an experimental lattice constant~\cite{lattice_constants}, ${a_\textrm{fcc}= 3.5906}$~\AA, 
that corresponds to Invar alloy with the nickel concentration 35 at. \%. 
To obtain on-site energies, $\epsilon^i_{m\s}$, a supercell containing 15 Fe atoms and 1 Ni atom was constructed. 
The Hamiltonian of the supercell is than written in the Wannier function basis set and
the difference between iron and nickel on-site energies
is defined as a difference between centers of gravity of corresponding $d$ states,
$\epsilon^{Fe}_d - \epsilon^{Ni}_d = \int ( \rho^{Fe}_d(\eps) - \rho^{Ni}_d(\eps) ) \, \eps \, d \eps$,
and it is equal to 0.88~eV~\cite{diff_splitting}. 
This calculation was carried out for the nickel impurity in the iron host,
hence, it is obvious that $\epsilon^{Fe}_d$=0~eV and $\epsilon^{Ni}_d$=-0.88~eV.

The CPA+DMFT calculations were performed with the AMULET code~\cite{amulet}.
To solve an auxiliary impurity problem (eq.~(\ref{eq:G_imp})) arising in DMFT 
a segment version of the continuous time 
quantum Monte-Carlo (CT-QMC) method~\cite{Werner2006} was employed.
This implementation of CT-QMC algorithm does not have sign problem and 
it is faster than other quantum Monte-Carlo methods (for more details see review~[\onlinecite{Gull2011}]
and Ph.D. thesis of E.~Gull~\cite{PhDGull}).
At the same time, it treats density-density terms of the Coulomb interaction in eq.~(\ref{eq:Hcoul}),
neglecting spin-flips and pair-hopping contributions.

\begin{figure}[bth!]
  \centering
  \includegraphics[clip=true,width=0.45\textwidth]{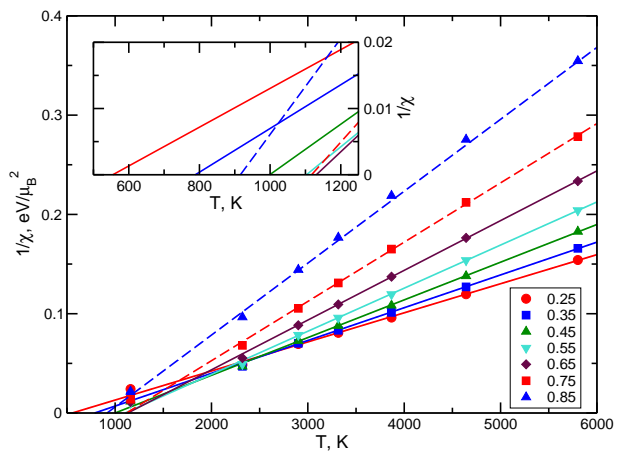}
  \caption{(Color online) Inverse of the uniform magnetic susceptibility, $\chi^{-1}(T)$,
           as function of temperature for various values of nickel concentration in \invar~(see colorcoding in figure). 
           Inset shows scaled up region around the Curie-Weiss temperature, $T_{CW}$.}
  \label{fig:chi}
\end{figure}

We used next values for the screened Coulomb interactions and Hund's exchange parameters, 
$U^{Fe}$=4~eV, $U^{Ni}$=6~eV and $J^{Fe}$=$J^{Ni}$=0.9~eV. 
These values are in good agreement with results of constrained local density approximation
calculations~\cite{cRPA} and used in earlier DFT+DMFT papers for elemental 
iron~\cite{Vekilova2015,Belozerov2014}.
Since the Coulomb interaction was already included in the density functional formalism,
and thus in the Hamiltonian, $H(\bk)$, 
we evaluate the double counting term according to the eq.~(\ref{eq:dc}).
%
%
%
%
We used DFT values for the total number of $d$ electrons, $N_d^{Fe}$=6.89 and
$N_d^{Ni}$=8.77, that leads to a fixed values of double counting.
%

Inverse of an uniform magnetic susceptibilities for various values of nickel concentrations 
are presented in Fig.~\ref{fig:chi}. 
Keeping in mind a smaller value of Coulomb interaction and lattice constant in
the paper of Vekilova \etal~\cite{Vekilova2015},
the obtained results are in good qualitative agreement
with the supercell DFT+DMFT data for Fe$_{0.75}$Ni$_{0.25}$.
The uniform magnetic susceptibility was calculated in a linear response manner by applying
a small external magnetic field, $B$, and evaluating magnetization of the compound, $m(T)$,
at given temperature, $\chi(T) = m(T) / B$. 
One can clearly see that for all Ni concentrations $\chi^{-1}$ has a linear behavior at high temperatures,
that allows one to extract Curie-Weiss temperatures, $T_{CW}$, and effective magnetic moments, $\mu_{eff}^2$,  
defined by the Curie-Weiss law,
\begin{equation}   \nonumber
  \chi_{CW}(T) = \frac{\mu_{eff}^2}{3 ( T - T_{CW} )}.
\end{equation}

\begin{figure}[bth!]
  \centering
  \includegraphics[clip=true,width=0.45\textwidth]{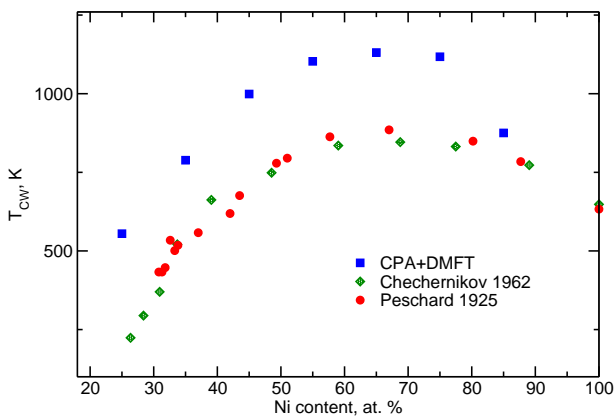}
  \caption{(Color online) CPA+DMFT and experimental values of Curie-Weiss temperature as function of Ni concentration.
           Experimental data are from Chechernikov~\cite{Chechernikov62} and 
           Peschard~\cite{Peschard25}.}
  \label{fig:Tc}
\end{figure}

The CPA+DMFT Curie-Weiss temperatures 
together with the experimental data~\cite{Chechernikov62,Peschard25,Swartzendruber91} are shown in Fig.~\ref{fig:Tc}.
The extracted $T_{CW}$ follow very well its experimental counterparts.
$T_{CW}$ is risen with increasing of nickel concentrations, than it has a maximum around 65 at. \% of Ni,
and finally decreases at higher Ni percentage.
In general, the theoretical Curie-Weiss temperatures are overestimated with respect to experimental values
by factor, $T_{CW}^{Th}/T_{CW}^{Exp} \approx$ 1.4 (except for endpoints).
This overestimation of the Curie-Weiss temperature is known in the DFT+DMFT 
calculations for various compounds~\cite{Lichtenstein2001,Leonov2011,Belozerov2011,Igoshev2013} and
it comes from two sources. 
The first source is a local nature of the DMFT approximation which leads
to a $\bk$-independent self-energy, and thus, neglecting spacial correlation effects that are important
in a vicinity of the magnetic transition.
Since the lattice is fixed the effect of neglecting spacial long range correlations 
on the uniform susceptibility is approximately
the same for various alloy concentrations.
The second reason of overestimation originates from
a density-density form of the Coulomb interaction in a segment version of CT-QMC algorithm~\cite{Sakai2006,Belozerov2012}.
The better agreement for the Curie temperature at high nickel content is attributed with lesser
importance of spin-flip and pair hopping terms in an almost filled $d$ shell~\cite{Antipov2012}.
Therefore, keeping in mind the above mentioned arguments, the theoretical CPA+DMFT Curie-Weiss temperatures
agree very well with the experimental data and, what is more important, they are changed with nickel
content in exactly the same way as in experiment.

\begin{figure}[bth!]
  \centering
  \includegraphics[clip=true,width=0.45\textwidth]{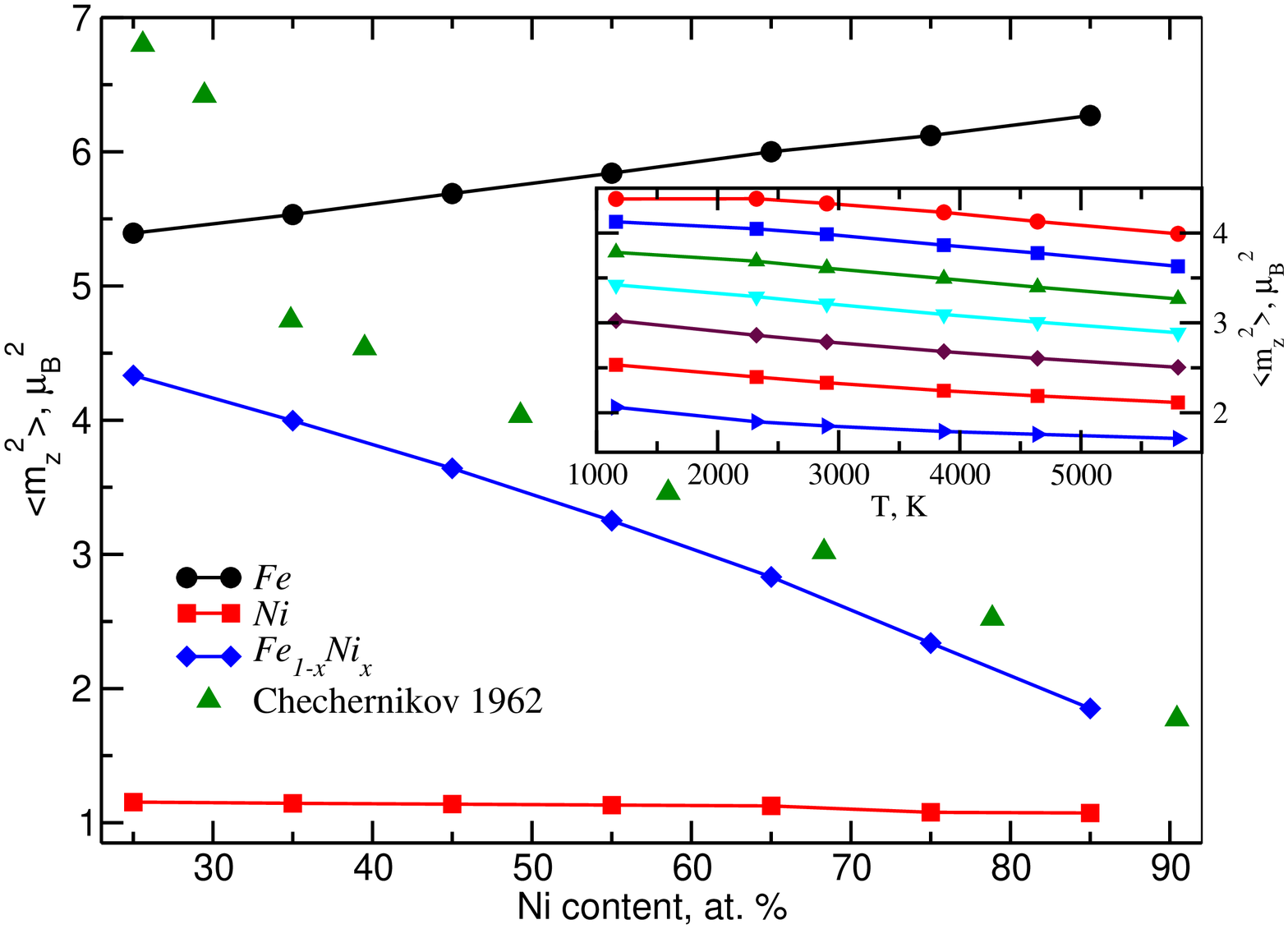}
  \caption{(Color online) Calculated at $T$=2900~K local squared magnetic moments, $\mz2$, for Fe, Ni and \invar~alloy
           as function of Ni concentration (for color-coding see figure).
           Experimental effective magnetic moments extracted from paramagnetic susceptibility~\cite{Chechernikov62}
           are shown by green triangles.
           Inset shows a temperature dependence of alloy's $\mz2$ for various nickel concentrations (color-coding is the same as in Fig.~\ref{fig:chi}).}
  \label{fig:mz2_vs_x}
\end{figure}

Local squared magnetic moments for iron, nickel and their weighted sum are presented in Fig.~\ref{fig:mz2_vs_x}.
The local squared magnetic moment is defined as
\begin{equation}    \nonumber
  \mz2^i = \langle \sum_{mm'} ( \hat{n}^i_{m\uparrow} - \hat{n}^i_{m\downarrow} )
                              ( \hat{n}^i_{m'\uparrow} - \hat{n}^i_{m'\downarrow} )  \rangle,
\end{equation}
where $i$=(Fe,Ni) and the corresponding quantity for the alloy is $\mz2 = (1-x)\mz2^{Fe} + x \mz2^{Ni}$.
Local squared magnetic moment of nickel is slowly decreasing with Ni concentration from 1.15 to 1.07 while
iron's function  is almost linearly increasing with values from 5.39 to 6.27.
The resulting alloy's $\mz2$ is quadratically decreasing and it follows 
the experimental effective magnetic moments extracted from paramagnetic susceptibility~\cite{Chechernikov62}.
The large discrepancy is observed at nickel concentrations below 30 at. \% and it is attributed to 
the experimentally observed coexistence of face centered cubic and 
body centered cubic phase~\cite{McKeehan1923} 
not presented in the calculations.
In spite of the fact that the local magnetic moments were calculated in paramagnetic phase, 
their concentration behavior corresponds to the atomically resolved 
experimental data~\cite{Shull1955,Cable1973,Glaubitz2011}
obtained for the magnetically ordered alloys and to the earlier DFT+DMFT results~\cite{Minar2005}.
Inset of Fig.~\ref{fig:mz2_vs_x} shows a temperature dependence of the local squared magnetic moment of
\invar~alloy in the concentration range $x$=[0.25...0.85]. For all studied concentrations it is decreased 
with temperature as in a pure fcc-Fe~\cite{Igoshev2013}.

\begin{figure}[b]
  \centering
  \includegraphics[clip=true,width=0.45\textwidth]{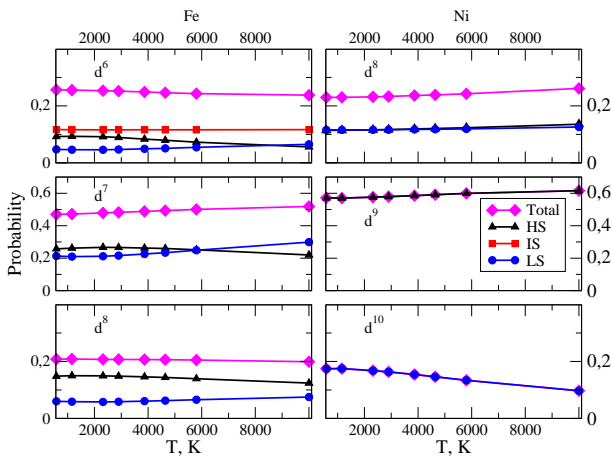}
  \caption{(Color online) Probabilities of different ionic configurations and spin states
           for Fe and Ni in Fe$_{0.75}$Ni$_{0.25}$ alloys as function of temperature.
           Weight of the ionic configuration (total) is shown by magenta diamonds, while high- (HS), 
           intermediate- (IS) and low- (LS) spin states are presented by black triangles,
           red squares and blue circles, respectively.}
  \label{fig:prob_Fe_0.75Ni_0.25}
\end{figure}

\begin{figure}[tb!]
  \centering
  \includegraphics[clip=true,width=0.45\textwidth]{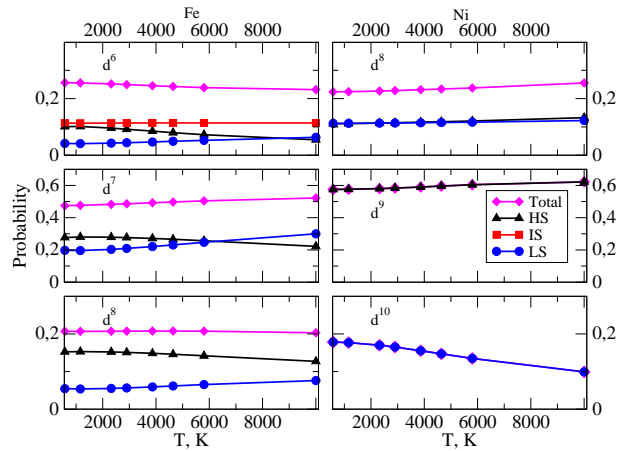}
  \caption{(Color online) Probabilities of different ionic configurations and spin states
           for Fe and Ni in Fe$_{0.65}$Ni$_{0.35}$ alloys as function of temperature.
           Color-coding as in Fig.~\ref{fig:prob_Fe_0.75Ni_0.25}.}
  \label{fig:prob_Fe_0.65Ni_0.35}
\end{figure}

\begin{figure}[tb!]
  \centering
  \includegraphics[clip=true,width=0.45\textwidth]{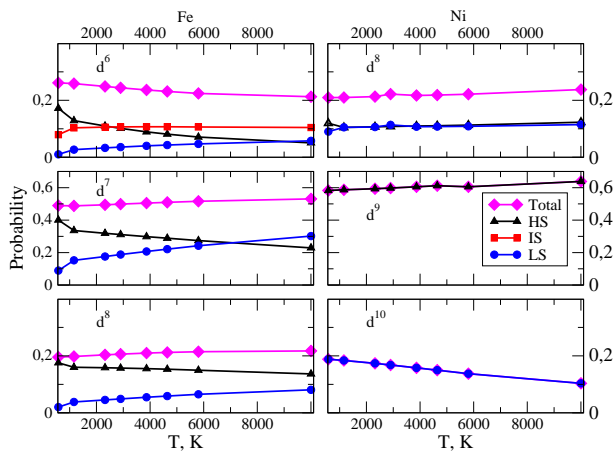}
  \caption{(Color online) Probabilities of different ionic configurations and spin states
           for Fe and Ni in Fe$_{0.35}$Ni$_{0.65}$ alloys as function of temperature.
           Color-coding as in Fig.~\ref{fig:prob_Fe_0.75Ni_0.25}.}
  \label{fig:prob_Fe_0.35Ni_0.65}
\end{figure}

\begin{figure}[htb!]
  \centering
  \includegraphics[clip=true,width=0.45\textwidth]{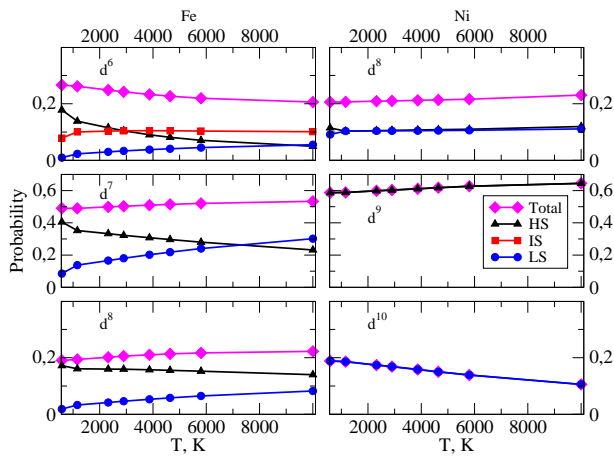}
  \caption{(Color online) Probabilities of different ionic configurations and spin states
           for Fe and Ni in Fe$_{0.25}$Ni$_{0.75}$ alloys as function of temperature.
           Color-coding as in Fig.~\ref{fig:prob_Fe_0.75Ni_0.25}.}
  \label{fig:prob_Fe_0.25Ni_0.75}
\end{figure}

To analyze in more details a temperature and concentration dependence of the magnetic properties
of constituents in \invar~we present distributions of ionic weights and corresponding spin configurations
in Figs.~\ref{fig:prob_Fe_0.75Ni_0.25}-\ref{fig:prob_Fe_0.25Ni_0.75}.
In the right panel of these figures the $d^8$, $d^9$ and $d^{10}$ ionic configurations
of nickel are shown by magenta diamonds. For all temperature and concentration ranges
the high-spin ($d^{5\uparrow}d^{4\downarrow}$) state of $d^9$ ionic configuration has the largest weight 
like in the pure nickel~\cite{Kolorenc2012}.
The weight of $d^8$ contribution is much smaller and it is
equally distributed between high-spin ($d^{5\uparrow}d^{3\downarrow}$) and low-spin ($d^{4\uparrow}d^{4\downarrow}$) states.
The $d^8$ and $d^9$ contributions are monotonically increasing with temperature.
The low-spin state of $d^{10}$ configuration is even smaller and it is decreasing with temperature.
All these states are almost independent of nickel percentage in \invar~and 
sum of the $d^8$ and $d^{10}$ total weights is about two times smaller than weight of $d^9$ ionic state. 
At the same time, the occupation of $d$ manifold in nickel is almost independent of temperature and concentration
in all CPA+DMFT calculations and it equals to 8.9$\pm$0.07 electrons, which is comparable with the DFT value of 8.77.

Situation with the states of iron presented in the left panel of Figs.~\ref{fig:prob_Fe_0.75Ni_0.25}-\ref{fig:prob_Fe_0.25Ni_0.75}
is more interesting.
The major contribution is formed by the $d^7$ ionic state but now satellite's total weights are comparable and
the sum of $d^6$ and $d^8$ ionic weights is approximately of $d^7$ weight.
For the $d^8$ ionic configuration the weight of the low-spin ($d^{4\uparrow}d^{4\downarrow}$) state is smaller than
the weight of high-spin ($d^{5\uparrow}d^{3\downarrow}$). The later is decreasing with temperature while the former is increasing
but they are not crossed in the temperature range of investigation.
This qualitative behavior of low- and high-spin states is kept for all nickel concentrations.
In the case of $d^7$ ionic configuration the weight of the high-spin ($d^{5\uparrow}d^{2\downarrow}$) state is increasing 
with the temperature, while the low-spin ($d^{4\uparrow}d^{3\downarrow}$) state is decreasing and
these spin configurations are intersecting around 6000~K with a shift to higher temperatures at nickel-rich alloy.

The situation with the $d^6$ ionic configuration is even more complicated because
an intermediate-spin ($d^{4\uparrow}d^{2\downarrow}$) state can be realized
besides the high- ($d^{5\uparrow}d^{1\downarrow}$) and low-spin ($d^{3\uparrow}d^{3\downarrow}$).
Exactly this intermediate-spin state dominates at high temperatures for all nickel concentrations and
it let the high-spin state be major at $\approx$ 50 at. \% and low temperatures.
Thus, in the iron-rich \invar~(see Figs.~\ref{fig:prob_Fe_0.75Ni_0.25} and~\ref{fig:prob_Fe_0.65Ni_0.35})
the high-spin state has more weight for $T <$ 2900~K. In fact, in the $d^6$ ionic state there are two points
of spin state alternations. At high temperatures the intermediate-spin is a major, the high-spin is a minor and
the low-spin state is in between. At $T\approx$ 7300~K the high-spin and the low-spin states are switched and then
at $T\approx$ 2900~K and for nickel concentrations $x>$ 0.45 the high-spin and the intermediate-spin states
are interchanged.
The 3$d$ Fe occupation is almost independent of temperature and concentration (like for nickel)
and it equals to 6.93$\pm$0.06 electrons, which is a bit higher than the DFT value of 6.89.

Albeit these spin state transitions in iron occur at the temperatures much above
the experimental Curie-Weiss one (keeping in mind that $T_{CW}^{Theor} = 1.4 T_{CW}^{Exp}$),
a high temperature local atomic physics of paramagnetic system can be traced back
to the magnetically ordered phase.
Thus these transitions can be regarded as a realization of two possible electronic states in $\gamma$-Fe
with the ferromagnetic high volume and the antiferromagnetic low volume states proposed by Weiss~\cite{Weiss1963}.
One can also think of this rich local physics as of  a variety of magnetically ordered
phases at low temperatures found in the conventional DFT calculations as function of volume 
(see Ref.~\onlinecite{Abrikosov2007} and references therein).


\section{Conclusion\label{sec:conclusion}}

We formulated and implemented in computer codes the CPA+DMFT method which combines a coherent potential approximation and dynamical
mean field theory in one technique and treats a substitutional disorder and strong electron-electron
correlations on equal footing.
Then this CPA+DMFT method were applied to study the magnetic properties of \invar~alloy
above the Curie temperature.
The calculated inverse of the uniform magnetic susceptibilities show a linear behavior at high temperatures and
extracted Curie-Weiss temperatures follow the experimental values as function of nickel concentration.
The individual local squared magnetic moments for iron and nickel are in good agreement with
experimental data.
The analysis of the contributions to the ionic and spin configurations shows the alternating of
the high-spin state and, intermediate- and low-spin states, as function of temperature that
is in agreement with the two state theory of Weiss~\cite{Weiss1963}.
These transitions can be regarded
as a high temperature precursor of a multiple magnetic orders at low temperatures found in the
density functional calculations.
Since they are strongly overestimated the further CPA+DMFT studies for different volumes and
interaction parameters are of great interests.

\section{Acknowledgments}

The authors are grateful to Yu.N. Gornostyrev, A.S. Belozerov and S.L. Skornyakov for useful discussions.
The study was supported by the Russian Scientific Foundation through the project 14-12-00473.


\begin{thebibliography}{10}
\bibitem{diff_splitting}
  Calculations for the 25 at. \% of nickel give the value of the on-site splitting between
  Fe and Ni equal to 0.97~eV which modifies slightly numerics and does not change 
  the main results qualitatively.
\end{thebibliography}
\end{document}